\documentclass[%
preprint,
 amsmath,amssymb,
 aps,
prl,
]{revtex4-1}

\usepackage{graphicx}
\usepackage{dcolumn}
\usepackage{bm}
\usepackage{hyperref}

\begin{document}

\preprint{APS/123-QED}

\title{Non-Ergodicity of Nose-Hoover chain thermostat in computationally achievable time}

\author{Puneet Kumar Patra}
\affiliation{%
 Advanced Technology Development Center, Indian Institute of Technology Kharagpur, West Bengal, India 721302 
}%
\author{Baidurya Bhattacharya}%
 \email{baidurya@civil.iitkgp.ernet.in}
\affiliation{%
Department of Civil Engineering, Indian Institute of Technology Kharagpur, West Bengal, India 721302 
}%

\begin{abstract}
The widely used Nose-Hoover chain (NHC) thermostat in molecular dynamics simulations is generally believed to impart the canonical distribution as well as quasi- (i.e., space filling) ergodicity on the thermostatted physical system (PS). Working with the standard single harmonic oscillator, we prove analytically that the two chain Nose-Hoover thermostat with  unequal thermostat masses approach the standard Nose-Hoover dynamics and hence the PS loses its canonical and quasi-ergodic nature. We also show through numerical simulations over substantially long times that for certain Poincare sections, for both the equal and unequal thermostat mass cases, the bivariate distribution function of position and momentum (${x,p}$) and of reservoir degrees of freedom ($\xi,\eta$) lose their Gaussian nature. Further, the 4-dimensional $x-p-\xi-\eta$ extended phase space exhibits two holes of non-zero measure. The NHC thermostat therefore does not generate the canonical distribution or preserve quasi-ergodicity for the PS.

\begin{description}
\item[PACS numbers]
05.10.-a, 05.45.Pq
\end{description}
\end{abstract}

\maketitle


Ergodicity of dynamics is a pre-requisite for obtaining statistical-mechanical properties of a system from a single dynamical trajectory, generated, for example, through molecular dynamics simulations \cite{ref18}. The ergodic hypothesis in this context may be stated as: given sufficiently long time, a single phase space trajectory of the system must visit all regions of the accessible phase space with the same relative frequency as in the phase space distribution. For a physical system, $S$, in equilibrium and in contact with a reservoir at constant temperature, $T$, the equations of motion must result in a trajectory consistent with the canonical distribution ${f\left(\mathbf{x,p}\right) \propto exp\left[-\beta E\left(\mathbf{x,p}\right)\right]}$. Here, ${\beta = \left(k_{B}T\right)^{-1}}$ where ${k_B}$ is the Boltzmann constant, and $E$ is the instantaneous energy of $S$. In this case, ergodicity can also be interpreted as space-filling dynamics (quasi-ergodicity).
Amongst the several temperature control algorithms available (velocity rescaling \cite{Woodcock_rescale_thermostat,Hoover_GIK_thermostat,Evans_GIK_thermostat_2, Evans_GIK}, deterministic \cite{Nose_Originial, Nose_Hoover,MKT_Thermostat,Hamilton_Modified_NH, Winkler_Thermostat,HH_thermostat,Bond_Leimkuhler_thermostat, BT_thermostat,Samoletov_Config_thermostat,PB_thermostat} and  stochastic \cite{Andersen_thermostat, Grest_stochastic_thermostat, Langevin_thermostat}), the deterministic thermostats possess the appeal of having autonomous and time reversible dynamics. However, our current understanding on ergodic characteristics of most deterministic thermostats remains primitive. For example unlike previously thought, not all two parameter thermostats are ergodic \cite{Hoover_mutifractal_dissipation_coexistence}. Possibly the simplest and most commonly used deterministic thermostat, the Nose-Hoover (NH) thermostat \cite{Nose_Hoover,Hoover_csm_book} also suffers from poor ergodicity in systems with small degrees of freedom \cite{Legoll_NH_non_ergodicity, Kusnezov_thermostat_scheme, Kusnezov_thermostat_scheme_v2, Hoover_mutifractal_dissipation_coexistence, Posch_Hoover_Vesely_NH_non_ergodicity}. If and only if the extended system (physical system + reservoir) is ergodic with respect to the invariant measure of system dynamics, $S$ has a canonical distribution \cite{Cho_ergodicity,Nose_MD_thermostat_review} with its dynamics being phase-space filling. 

The Nose-Hoover chain (NHC) and Hoover-Holian thermostats have generally been considered to resolve the ergodicity issues of NH thermostat in equilibrium \cite{Hoover_mutifractal_dissipation_coexistence, ref18, Watanabe_NH_ergodicity, HH_thermostat}. In this work, we focus only on NHC thermostat. NHC controls the fluctuations of thermostat ${\eta}$ by coupling it with a new thermostat variable, ${\xi}$ \cite{MKT_Thermostat,ref18}. Fluctuations of the second thermostat (${\xi}$) can likewise be controlled with a third and so on thus forming a chain. The dynamics with two chains NHC thermostat for a single harmonic oscillator with unit mass and spring constant becomes:
\begin{equation}
\begin{array}{cc}
\dot{x} = p; &
\dot{p} = -x - \dfrac{\eta p}{Q_{\eta}}\\
\dot{\eta} = p^2 - k_{B}T - \dfrac{\eta\xi}{Q_{\xi}}; &
\dot{\xi} = \dfrac{\eta^2}{Q_{\eta}} - k_{B}T 
\end{array}
\label{eq:six}
\end{equation}
However, no conclusive proof of its ergodicity has been put forward so far \cite{Watanabe_NH_ergodicity} and all previous attempts have involved proving that the marginal distributions of position ($x$) and velocity ($p$) \cite{MKT_Thermostat, Tuckerman_JPCB_2000} are Gaussian. 

In this work, we show conclusive evidence that (i) NHC dynamics is not ergodic (ii) the physical system does not follow the canonical distribution at every Poincare section, and (iii) when the difference in thermostat masses is large, the dynamics due to NHC reduces to NH dynamics. We also conjecture how previous studies might have missed these points. We limit the scope of this work to two chain NHC thermostat.

It can be easily shown that the invariant measure of NHC phase space is:
\begin{equation}
d\mu = \dfrac{1}{Z}e^{-\frac{\beta }{2}x^2} e^{-\frac{\beta }{2} p^2} e^{-\frac{\beta }{2Q_\eta}\eta ^{2}} e^{-\frac{\beta }{2Q_\xi} \xi^{2}} dx dp d\eta d\xi
\label{eq:measure_NHC}
\end{equation}
where, $Z$ is the normalizing constant. Due to the statistical independence of all the variables, their marginal densities are Gaussian. Further, the conditional density function, $f(x,p|\eta=\eta_0,\xi=\xi_0)$, at fixed values of $\xi=\xi_0$ and $\eta = \eta_0$ is uncorrelated bivariate normal:
\begin{equation}
f(x,p|\eta=\eta_0,\xi=\xi_0) = \dfrac{1}{Z'}e^{-\beta x^2/2}e^{-\beta p^2/2}
\label{eq:bivar_cond_dist}
\end{equation}
 (\ref{eq:bivar_cond_dist}) must hold true for all Poincare sections for NHC to sample from canonical distribution and any deviation of joint probability distribution function (JPDF) from bivariate Gaussian indicates non-canonical nature of the dynamics. If such non-canonical dynamics exist, then the overall dynamics must be non-ergodic with holes present in the phase-space. Similar argument holds true for conditional distribution of $f(\eta,\xi|x=x_0,p=p_0)$. Similar argument holds true for univariate conditional distribution function: $f\left(x|p=p_{0},\eta=\eta_0,\xi=\xi_0\right)$.

We simulated the extended system involving the harmonic oscillator (\ref{eq:six}) at $\beta=1$ using four thermostat mass pairs: $Q_\eta,Q_\xi= (1,1), (10,0.1), (50,0.02)$ and $(100,0.01)$. Since the fluctuations of reservoir $\eta$ are controlled by the second reservoir $\xi$, effective thermostatting of $\eta$  can occur only if $Q_\xi \ll Q_\eta$. Equations of motion (\ref{eq:six}) were integrated using Runge-Kutta-Fehlberg method and run for 200 billion time steps each of 0.001. Various initial conditions were chosen. The canonical nature of the physical system, or equivalently, the ergodic nature of the extended system, was investigated on various Poincare sections of the 4-D phase space through joint moments, Kullback-Leibler distance, Hellinger distance and presence of holes of non-zero measure.

\begin{figure*}
\includegraphics[scale=0.45]{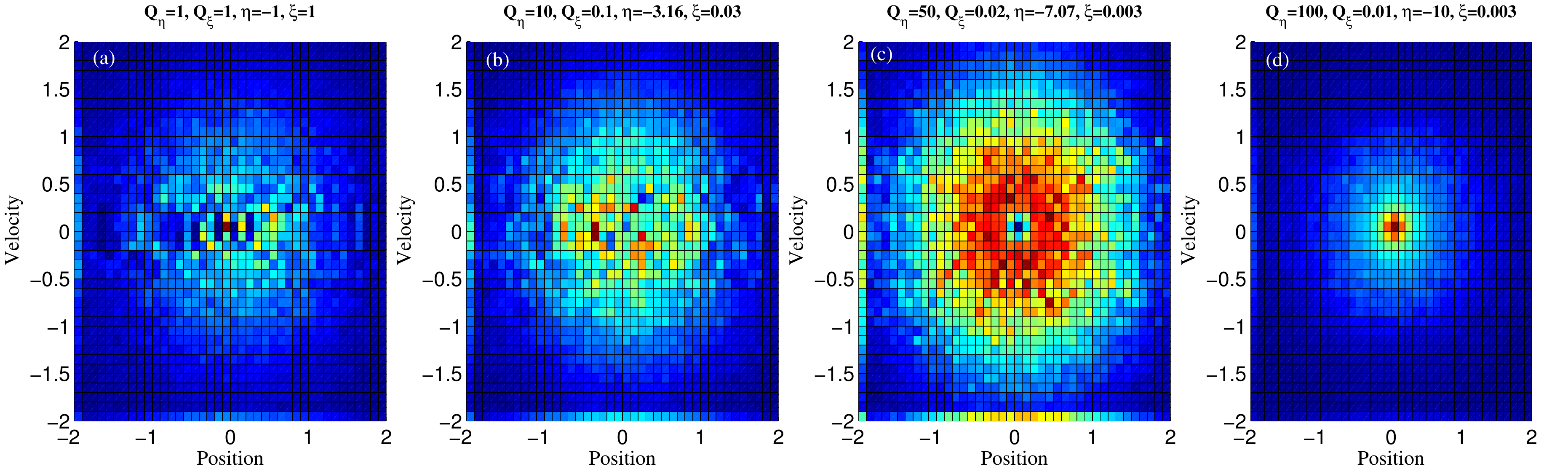}
\caption{\label{fig:plot1} Joint probability density (JPDF) plot of $x-p$ corresponding to different Poincare sections for four different pairs of thermostat masses with initial conditions as $x=1.1,p=1.1,\eta=0,\xi=0$. For each case we can observe that there is a deviation from normal distribution near origin.}
\end{figure*}

\begin{figure*}
\includegraphics[scale=0.45]{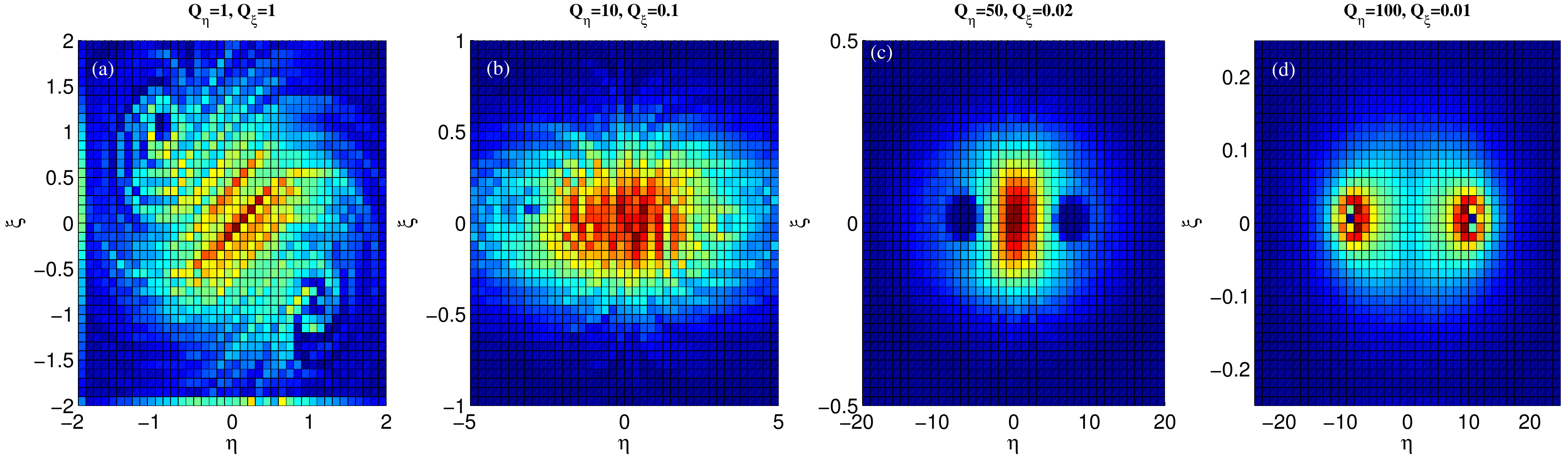}
\caption{\label{fig:plot3} ${\eta-\xi}$ joint probability density for Poincare section at $x=0,p=0$. The figures indicate presence of two holes in the system dynamics.}
\end{figure*}

When initial conditions were chosen far from fixed points (given by $x=0,p=0,\eta=\pm \sqrt{Q_{\eta}}, \xi = \mp Q_{\xi}/\sqrt{Q_{\eta}}$), the trajectories never came close to the fixed point during the simulation duration. Four such Poincare sections initialized at $x=1.1,p=1.1,\eta=0, \xi=0$ are shown in Figure \ref{fig:plot1}. In each case there is a significant deviation from bivariate normal near the origin ($x=0,p=0$): the JPDF is in fact zero in Figures \ref{fig:plot1}(a) and (c). This deviation would be missed if instead, the JPDF of $x-p$ is obtained by projecting the dynamics from the 4d space onto the $x-p$ plane. The deviation from Gaussian becomes even more apparant when one studies the Poincare section defined at $x=0$ and $p=0$. In fact, one can conclude just by looking at the JPDFs of $\eta,\xi$(see Figure \ref{fig:plot3}) that they are not bivariate Gaussian. 

We checked the convergence in distribution of the JPDF ($f_1$) of $x,p$ (RHS of \ref{eq:bivar_cond_dist}) to the uncorrelated standard bivariate normal ($f_2$) with the help of Hellinger \cite{Hellinger_distance} and symmetrical form of Kullback-Leibler (KL) \citep{KLD_Paper} distances:
\begin{equation}
D_{H} \left(f_{1}||f_{2} \right)= \frac{1}{\sqrt{2}} \sqrt{\sum_{i,j} \left( \sqrt{f_{1}\left(i,j\right)} - \sqrt{f_{2}\left(i,j\right)}\right)^2}
\label{eq:eleven}
\end{equation}

\begin{equation}
D_{KL} \left(f_{1}||f_{2} \right)= \sum_{i,j} \left[ f_1\left( i,j \right)\text{ln}\dfrac{f_1\left( i,j \right)}{f_2\left( i,j \right)}\\ + f_2\left( i,j \right)\text{ln}\dfrac{f_2\left( i,j \right)}{f_1\left( i,j \right)} \right]
\label{eq:kld}
\end{equation}

\begin{figure} 
\includegraphics[scale=0.50]{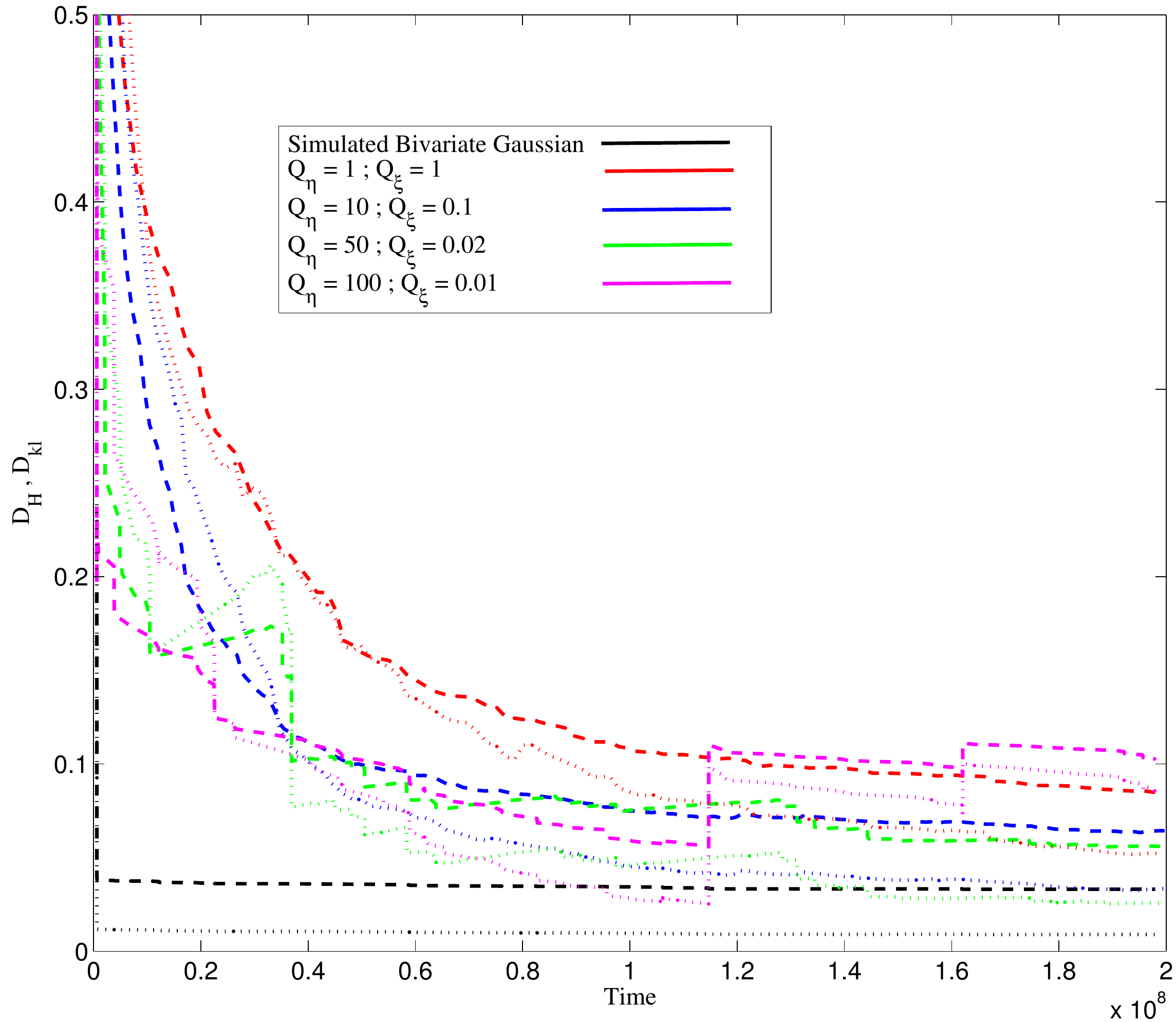}
\caption{\label{fig:plot2} Convergence of Hellinger (thick dotted lines) and modified Kullback-Leibler distances (fine dotted lines) for various cases. Bins of size $\Delta x = 0.1$ and $\Delta p = 0.1$ are used. None of the distributions converges to the standard uncorrelated bivariate normal}
\end{figure}

The evolution of the Hellinger and KL distances between the uncorrelated standard bivariate normal and the JPDF of ($x,p$) at each of the Poincare sections of Figure \ref{fig:plot1} is shown in Figure \ref{fig:plot2}. Despite long simulations, the distributions do not converge. We also looked at the first 6 even joint moments of $x,p$:$\langle x^{2n}p^{2n}\rangle$ (n varying from 1 to 6) and found that deviations were as high as 43$\%$, 75$\%$, 40$\%$ and 226$\%$ respectively in the four cases from those of standard uncorrelated bivariate normal. Previous studies have used marginal distributions as a basis to argue that NHC generates canonical dynamics. We too found that non-Gaussian features are overlooked if one looks only at the marginal distributions, or JPDFs obtained through projections on a plane. Examples where marginal Gaussian variables do not produce the joint Gaussian distribution may be found in standard probability texts(for example \cite{Feller_book}). 

\begin{figure*}
\includegraphics[scale=0.45]{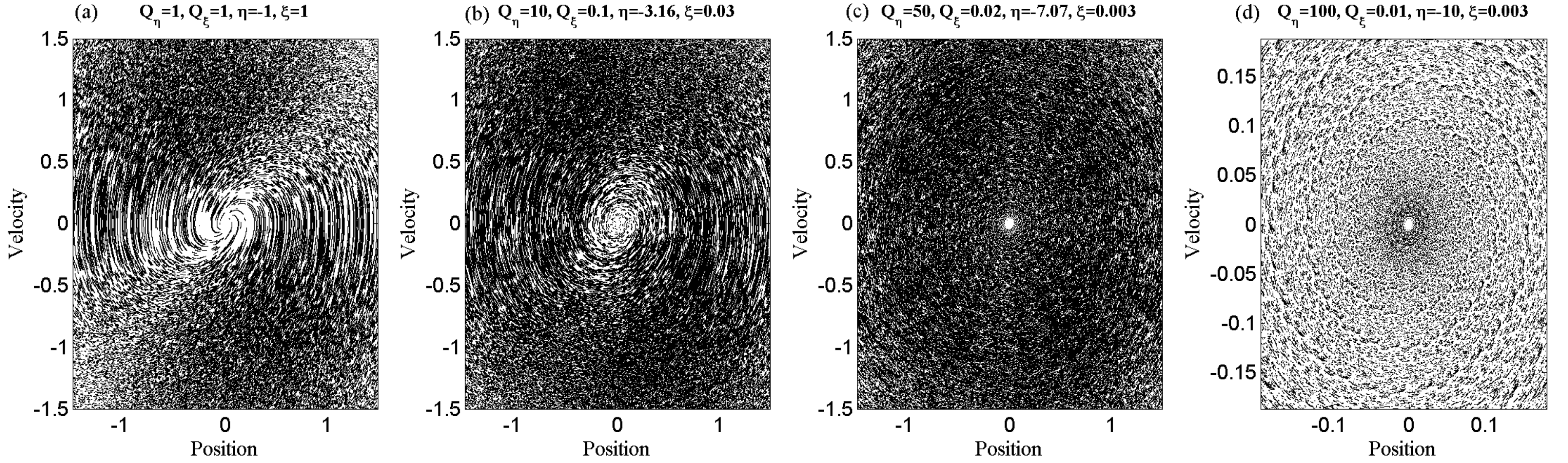}
\caption{\label{fig:plot4} Phase space plot of $x-p$ corresponding to different Poincare sections for four different pairs of thermostat masses. Existence of unoccupied regions near origin for all the cases suggests that holes are present in phase space.}
\end{figure*}

\begin{figure*}
\includegraphics[scale=0.45]{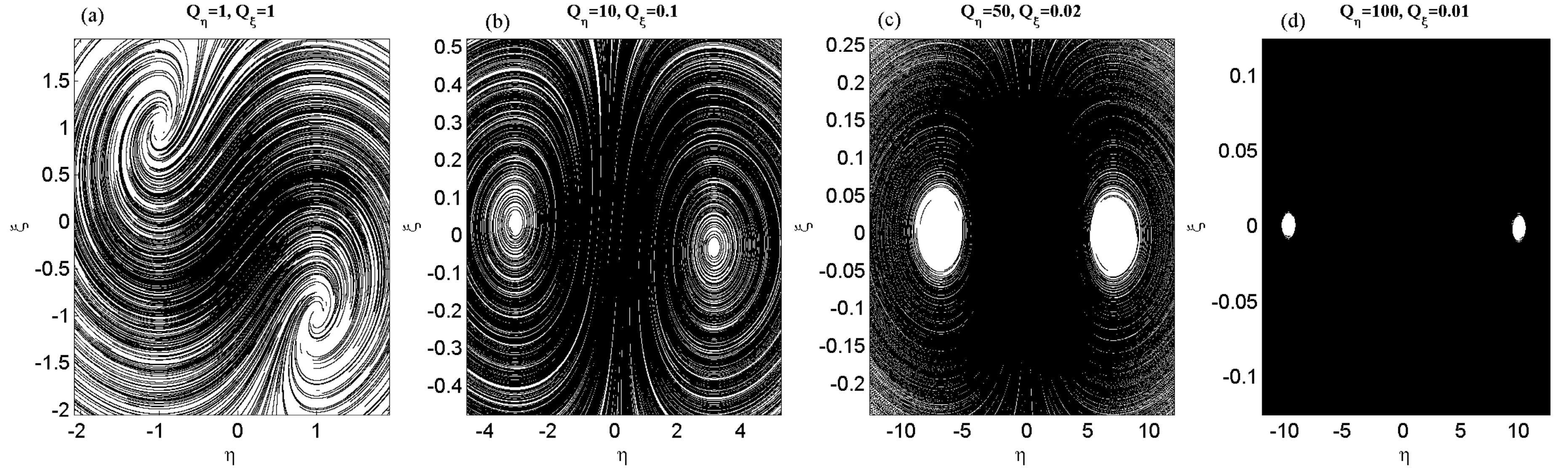}
\caption{\label{fig:plot5} Phase space plot of $\eta-\xi$ corresponding to different Poincare sections for four different pairs of thermostat masses. Two holes can be easily seen from the figure.}
\end{figure*}

\begin{figure*}
\includegraphics[scale=0.45]{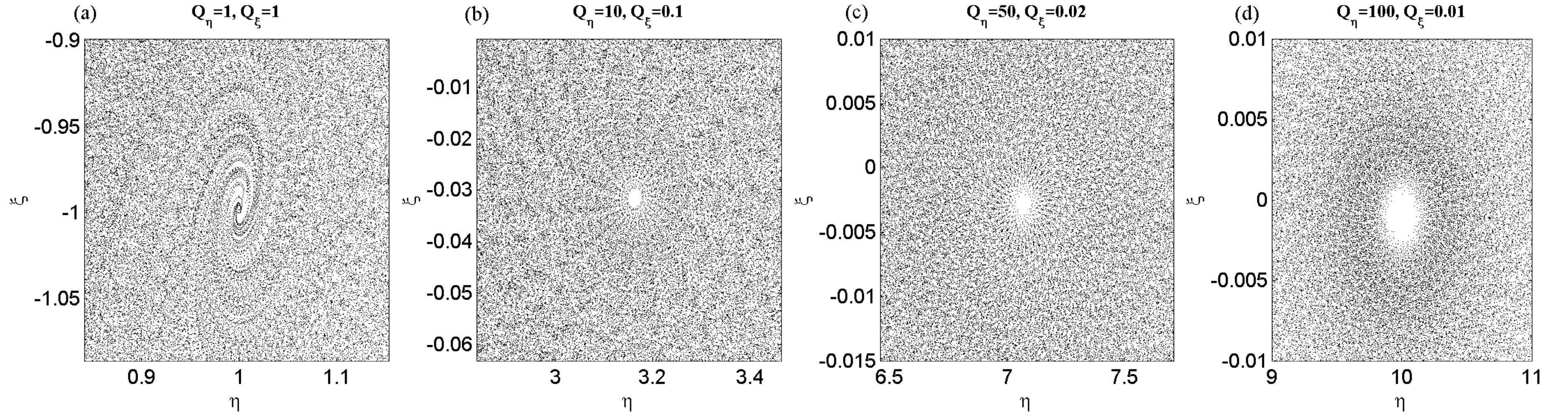}
\caption{\label{fig:plot6} Non-zero measure holes in each of the four cases using 20 billion time steps each of 0.01. In each of the figures, $-0.1 \le x \le 0.1; -0.1 \le p \le 0.1$.}
\end{figure*}

The non-ergodic nature of the dynamics is further revealed by the phase-space plots in Figures \ref{fig:plot4} and \ref{fig:plot5} (corresponding to Figures \ref{fig:plot1} and \ref{fig:plot3}, respectively). Holes are clearly present in the phase space. As the difference of the thermostat masses decreases, the holes rotate in the plane. Due to the complexity of the dynamics in 4-dimensional space, attempts to find stable periodic orbits were unsuccessful. We therefore take an alternate route to confirm that the holes are of non-zero measure. We focused around one of the possible locations of holes because of anti-symmetry of Figure \ref{fig:plot5}. Instead of working with Poincare sections, we took a splice of much larger width in $x-p$. We looked at the region ${-\lambda \le x \le \lambda}$; ${-\lambda \le p \le \lambda}$ and kept increasing $\lambda$ until ${\eta-\xi}$ plot showed no existence of sparesely populated region. If the unoccupied regions in Figures \ref{fig:plot4} and \ref{fig:plot5} were limited to just one hyperplane, i.e. $\lambda \approx 0$ when sparsely populated regions disappear, then the measure of the holes present in the system would be zero and the dynamics would still be ergodic. The corresponding JPDFs would have been uncorrelated bivariate standard normal. Figure \ref{fig:plot6} shows the presence of very small unpopulated regions when ${\lambda = 0.1}$. This confirms that there is a hole whose $x-p$ boundary is given by  ${|x| \approx 0.1}$ and ${|p| \approx 0.1}$ . Further, unlike other cases, the hole for the case of $Q_\eta=100,Q_\xi=0.01$ was found to be a through hole. The size of the hole progressively decreases and reaches a minimum value when both thermostat masses are set at 1 (but does not remain confined to one hyperplane and hence, is of non-zero measure). 

Theoretically, it is possible to understand why NHC shows through holes for large difference in masses. Using the transformation ${\eta'=\eta/Q_{\eta}}$ and ${\xi'=\xi/Q_{\xi}}$,  (\ref{eq:six}) can be written in terms of ${\eta'}$ and ${\xi'}$. Choosing ${x=r\sin\theta}$ and ${p=r\cos\theta}$, and rearranging in terms of ${\dot{r}}$ and ${\dot{\theta}}$, (\ref{eq:six}) can be rewritten as:
\begin{equation}
\begin{array}{cc}
\dot{r} = -\eta' r \cos^{2}\theta; &
\dot{\theta} = 1 + \eta'\sin\theta \cos\theta \\
\dot{\eta'} = \dfrac{1}{Q_\eta}\left(r^2\cos^2\theta - k_{B}T\right) - \eta' \xi'; &
\dot{\xi'} = \dfrac{\eta'^2 Q_\eta}{Q_{\xi}} - \dfrac{k_{B}T}{Q_{\xi}}
\end{array}
\label{eq:na}
\end{equation}
Under the scenario ${Q_{\xi} \le Q_{\eta}}$, the fluctuations of ${\xi'}$ occur at a much faster rate than other variables. As a result, ${\xi'}$ may be replaced by its average. Once the steady state is reached, ${\xi'}$ fluctuates around the mean and consequently, its average is zero. (\ref{eq:na}) can be written in terms of three variables:
\begin{equation}
\begin{array}{ccc}
\dot{r} = \eta' r \cos^{2}\theta; &
\dot{\theta} = 1 + \eta'\sin\theta \cos\theta; &
\dot{\eta'} = \dfrac{\left(r^2\cos^2\theta - k_{B}T\right)}{Q_\eta}
\end{array}
\label{eq:na1}
\end{equation}
(\ref{eq:na1}) is same as the standard Nose-Hoover dynamics \cite{Watanabe_NH_ergodicity} and hence shows similar features (as well as problems) as the NH thermostat. The Hoover-Holian thermostat, on the other hand, does not show problems highlighted before.

\textit{To summarize}, we used numerical simulations to demonstrate that in the rather long time duration considered, Nose-Hoover chain thermostat is unable to generate a canonical distribution in certain Poincare sections, despite the overall projection being quite close to canonical. This occurs due to the presence of two 4-dimensional holes of non-zero measure. Merely showing the Gaussian nature of marginal distributions of position and velocity, as has been done in the past, is not sufficient to prove ergodicity. Since the trajectories of $x,p,\eta$ and $\xi$ each constitutes a stochastic process, it can also be argued that the EOMs (\ref{eq:six}) do not support the assumption of ergodicity (\ref{eq:measure_NHC}) which implies each of the four processes is stationary Gaussian. For if $p$ is a Gaussian process, so is its derivative $\dot{p}$, and if $p$ is stationary Gaussian, it is also independent of $\dot{p}$ at the same instant - both of which are violated by (\ref{eq:six}) (and likewise for $\eta$ and $\xi$). We however do not probe this stochastic process angle further in this letter. 

\begin{acknowledgements}
We wish to thank Prof. William G. Hoover for his several useful comments that helped in improving the quality of this document.
\end{acknowledgements}

\bibliography{apssamp}
\end{document}